# Topology Control Schema for Better QoS in Hybrid RF/FSO Mesh Networks

Osama Awwad*, Ala Al-Fuqaha*, Bilal Khan†, Ghassen Ben Brahim‡

*Computer Science Department, Western Michigan University

{oaawwad, alfuqaha}@cs.wmich.edu

†John Jay College, City University of New York.

bkhan@jjay.cuny.edu

‡Integrated Defense Systems, Boeing Company

ghassen.benbrahim@boing.com

*Abstract*— The practical limitations and challenges of radio frequency (RF) based communication networks have become increasingly apparent over the past decade, leading researchers to seek new hybrid communication approaches. One promising strategy that has been the subject of considerable interest is the augmentation of RF technology by Free Space Optics (FSO), using the strength of each communication technology to overcome the limitations of the other. In this article, we introduce a new scheme for controlling the topology in hybrid Radio-Frequency/Free Space Optics (RF/FSO) wireless mesh networks. Our scheme is based on adaptive adjustments to both transmission power (of RF and FSO transmitters) and the optical beam-width (of FSO transmitters) at individual nodes, with the objective of meeting specified Quality of Service (QoS) requirements, specifically end-to-end delay and throughput. We show how one can effectively encode the instantaneous objectives and constraints of the system as an instance of Integer Linear Programming (ILP). We demonstrate that the technique of Lagrangian Relaxation (LR), augmented with iterative repair heuristics, can be used to determine good (albeit sub-optimal) solutions for the ILP problem, making the approach feasible for mid-sized networks. We make the proposed scheme viable for large-scale networks in terms of number of nodes, number of transceivers, and number of source-destination pairs by solving the ILP problem using a Particle Swarm Optimization (PSO) implementation.

## I. INTRODUCTION

Most present day wireless networks are deployed using only radio frequency (RF) channels, since these provide efficient support for radial signal broadcasting by each of the network's constituent nodes. The disadvantages of RF communication are by now well-known, including bandwidth scarcity, lack of security, high interference, and high bit error rates. These limitations make providing scalable quality of service (QoS) support difficult, if not intractable.

Faced with such daunting obstacles to QoS in RF-only networks, the use of Free Space Optics (FSO) for wireless communications was proposed [6]. FSO technology enables the transmission of data using optical signals through free space (or air), and so has the potential to support higher link data rates than RF. The propagation of optical signals requires the use of specialized light sources [1], such as lasers (providing coherent light) and LEDs (providing non-coherent light). LEDs are better suited for wireless mesh networks, since they consume very little power, have wider beam-width compared with lasers, and overcome the safety issues that are a major concern with high-powered lasers. Recent developments in high-brightness LED technology has made it possible to deploy LED transmitters with a rate up to 2 Gbps, with transmission ranges of up to 104 miles [2]. In addition to high data rates, because FSO uses directed optical transmissions with adjustable channel beam-width, inter-FSO communication interference can be minimized. In addition, because FSO does not support radial broadcasting, it also provides some degree of security against malicious eavesdropping. The benefits of FSO do not come without a price, however. Most notable of these drawbacks is the need to maintain line of sight (LOS) between the transmitter and the receiver during the course of communication. In contrast, LOS is not a requirement for RF transmissions.

The complementary strengths and weaknesses of RF and FSO make them ideal technologies to hybridize. A hybrid approach that uses both RF and FSO has the potential to overcome the weakness of each of the individual channel types. Several frameworks have been proposed for such hybrid networks [3, 4, 5, 23, 25] and a few prototypes have been implemented [7, 8, 24].

The dynamic nature of RF and FSO channel characteristics makes topology control a major issue in hybrid RF/FSO networks. Questions of topology control have already been studied extensively in traditional RF-only networks. In the RF-only context, the objective is to adjust the "microscopic" node-level parameter (e.g. transmission power levels) so as to achieve some "macroscopic" network-level property such as connectivity, minimal interference, or specified QoS requirements [9, 10, 17]. The newer questions of topology control in hybrid RF/FSO networks are much more challenging, and by comparison, relatively few solutions have been developed; here we survey some of the notable prior achievements. In [11], Kashyap et al. proposed a joint topology control and routing framework wherein the RF links serve to provide instantaneous backup to traffic in hybrid RF/FSO networks whenever FSO links become degraded. In another paper [12], the same authors studied the ability to provide topology reconfiguration in response to changes in links capacities and traffic demands in RF/FSO networks. The authors proposed a heuristic for finding a topology configuration with the minimum packet dropping rate[1]. We are not aware of any previous work which has explored the potential for optimization introduced by considering dynamic adjustment of both the RF and FSO transmission power levels and the FSO beam-widths in RF/FSO networks. This is in fact, the subject of we address here.

We propose an adaptive topology control framework wherein each node can adjust its power level for both RF and FSO channels, while simultaneously adjusting the beam-width of FSO channels. The objective of such adjustments is a topology configuration that meets given QoS requirements specified in terms of end-to-end delay and throughput, all while minimizing the total power consumed for transmissions. More precisely, we assume that the *cost* of RF and FSO channels depends on current channel characteristics, while the topological *structure* is mandated by the desired QoS. The proposed system model seeks to determine structure given costs, and is flexible enough to be applied to any hybrid network consisting of both omnidirectional/directional communication channels[2] (of which RF/FSO networks are one very important special case).

---

[1] The authors took the packet dropping rate to include both link congestion and packet drops that occur in the transient states occurring during a topology reconfiguration.

The decision to use a hybrid model brings with it a unique set of challenges, stemming from the fact that nodes can choose between two *different* channels types—each with its own transmission characteristics. Because of this freedom, steps must be taken to prevent a relay node in a multi-hop connection from being "tempted" to behave selfishly by forwarding other nodes' packets using the less reliable of the two channel types, thereby avoiding the individual opportunity cost that would be incurred by a "fairer" choice of allocating bandwidth on a high quality link. We addressed this problem of selfish behavior in [22] by formulating node decisions within a hybrid RF/FSO network in a Bayesian game-theoretic model that is designed specifically to guarantee optimal cooperativeness.

The rest of this paper is organized as follows. In Section II, we describe the underlying system model of hybrid RF/FSO channels. In Section III, we define the adaptive topology control problem and develop an integer linear programming (ILP) formulation for the problem. In Section IV, the ILP problem is addressed via Lagrangian Relaxation (LR) with iterative repair heuristic. In section V, a Particle Swarm Optimization solution for large-scale hybrid RF/FSO networks is devised. In Section VI, we present the experiments of the topology control problem in hybrid RF/FSO and interpret their outcomes. Finally, in Section VII, we present overall conclusions and the future trajectory of our research efforts.

## II. PROBLEM DEFINITION

Adjusting the beam-width and the transmission power in RF/FSO MANETs has both benefits and disadvantages. A node with large beam-width or high transmission power usually has more nodes in the transmission range (i.e. higher node degree). This can help to reduce the average number of hops, thus minimizing the end-to-end delay. However, higher node degrees can lead to more channel contentions and an increase in the amount of interference. Analogously, when nodes have a lower number of peers in their transmission range, they have lower connectivity, which results in high average path length, and consequently high end-to-end delay. As such, there is an apparent trade-off; our objective is to construct a robust topology by minimizing the total transmission power, by adapting the beam-width and selecting different channels such that we meet joint throughput and end-to-end delay requirements.

---

[2] All that would be required is an updating of modulation techniques described in the system model subsection.

In our system model, we consider a wireless mesh network consisting of a set $V$ of hybrid RF/FSO nodes $|V|=N$, where each node is equipped with directional FSO transceivers and omnidirectional RF transceivers with limited transmission range. For *each* node $i \in V$, we maintain the following information:

- Location.
- Number of RF and FSO transceivers[3].

The set of transceivers at node $i$ is denoted $T_i$. For *each* transceiver $t$ in $T_i$, we maintain the following information:

- *C_MAX*: Maximum capacity,
- *S*: Sensitivity,
- *D*: Diameter,
- Max Beam-Width,
- Max Power level.

In our model, nodes can adjust their FSO transmission beam-width and their RF and FSO transmission powers. We discretize these choices by specifying a uniform

- *P*: Set of possible transmission powers,
- *Φ*: Set of possible beam openings.

---

[3] We assume the light source for FSO channels is non-coherent light LEDs, where every FSO transceiver operates at its own unique wavelength, and thus multiple FSO transceivers can send and receive data at the same time without interfering with each other.

The strength of RF and FSO signals suffers from different attenuation losses during propagation [1, 13]. In our study, we focus on the attenuations that occur due to geometrical loss. If node $i$ transmits with power $P_t$ then node $j$ receives the signal with power $P_r$ given by [1]:

$$P_r = P_t \left( \frac{D_t}{D_t + 100 d\theta} \right)^2,$$

where $D_t$ is the transceiver diameter, $d$ is the distance between node $i$ and node $j$, and $\theta$ is the beam divergence angle.

Each wireless channel has a computable Bit Error Rate (BER), which is the probability of the occurrence of an error during data transfer over that link. The relationship between the BER of a wireless channel and the received power level $P_r$ is a function of the modulation scheme. In RF channels, we consider the instantaneous channel BER based on the non-coherent binary orthogonal Frequency Shift Keying (FSK) modulation scheme, as given in [14, 15]:

$$BER = \frac{1}{2} erfc\left( \frac{-P_r}{2 P_{noise}} \right),$$

where $P_{noise}$ is the RF noise power. In FSO channels, we consider BER based on the On-Off Keying (OOK) modulation scheme, as given in [16]:

$$BER = \frac{1}{2} erfc\left( \frac{R.P_r}{2\sqrt{2 P_{noise}}} \right),$$

where $R$ is the photo detector responsivity, and $P_{noise}$ is the FSO noise power.

In summary, our assumptions in the system model described above include: (i) the network topology is a wireless mesh with directed links, (ii) at any given node, both RF and FSO transceivers may be present, and (iii) RF transceivers are omnidirectional, while FSO transceivers are directional.

## III. ILP FORMULATION

We model our optimization problem as an instance of Integer Linear Programming (ILP). The input includes a set of *N* nodes *V*, together with associated information about each node and its RF and FSO transceivers, as described in the previous section. Additionally, as input, we have the QoS requirements:

- *SD*: Set of requested source-destination connections.

    For each $(s,d) \in SD$, we have:

    a. $H_{(s,d)}$: Maximum delay

    b. $Th_{(s,d)}$: Minimum Throughput

Initially, we construct a *possible* network topology G = (V, E), and our objective is to adapt select links from this topology adaptively; the adaptation will correspond to solving a specific instance of ILP. The set of potential links which might arise within *G* are enumerated via a set of indicator variables $l_{i,j,t}^{p,\theta_t,\theta_r}$ where for each pair of nodes *i, j* (in *V*), each transmitter *t* (in $T_i$), each transmission power *p* (in *P*), each transmitter beam-width $\theta_t$ (in $\Phi$), and each receiver beam-width $\theta_r$ (in $\Phi$), we take $l_{i,j,t}^{p,\theta_t,\theta_r} = 1$ if there is a *link(i,j,t)* available from node *i* to node *j* using transceiver *t*; otherwise we take $l_{i,j,t}^{p,\theta_t,\theta_r} = 0$. Specifically, the *link(i,j,t)* is considered "available" whenever the following conditions are met:

1) Node *j* is inside the coverage area of transceiver *t* at node *i*. This can be verified easily by first calculating the transmitter *t*'s maximum range, and then determining the coverage area based on a sector template for FSO channels or the circular template for RF channels.

2) Additionally, whenever *t* is an FSO transmitter, then nodes *i* and *j* must be in line of sight of each other. This second condition does not apply when *t* is an RF channel, since we assume omnidirectional RF transmission.

3) For each tuple *i, j, p,* $\theta_t$, $\theta_r$ for which $l_{i,j,t}^{p,\theta_t,\theta_r} = 1$, we define:

    - $BER_{i,j,t}^{p,\theta_t,\theta_r}$, the bit error rate on *link (i,j,t)*.

- $B_{i,j,t}^{p,\theta_t,\theta_r}$, the bandwidth of *link(i,j,t)* computed according to the formula $B_{i,j,t}^{p,\theta_t,\theta_r} = \dfrac{B\max_{i,t}}{\sum l_{i,j,t}^{p,\theta_t,\theta_r}}$, where $B\max_{i,t}$ is the bandwidth of $t$ (in $T_i$).

Now, building on the above basic variables, we introduce additional indicator **variables** which implicitly form the basis of the space of solutions. For each node $i$ (in V), each transmitter $t$ (in $T_i$), and each transmission power $p$ (in P), we introduce the indicator variable $x_{i,t}^p$, defined as:

- $x_{i,t}^p = 1$ if transceiver $t$ at node $i$ is transmitting using power $p$); otherwise $x_{i,t}^p = 0$.

Additionally, for each pair of nodes $i, j$ (in V), each transmitter $t$ (in $T_i$), each transmitter beam-width $\theta_t$ (in $\Phi$), and each receiver beam-width $\theta_r$ (in $\Phi$), and each transmission power $p$ (in P), we introduce the indicator variable $g_{i,j,t}^{p,\theta_t,\theta_r}$, defined as:

- $g_{i,j,t}^{p,\theta_t,\theta_r} = 1$ if $l_{i,j,t}^{p,\theta_t,\theta_r}$ is selected in the constructed topology; otherwise $g_{i,j,t}^{p,\theta_t,\theta_r} = 0$.

Finally, for each pair of nodes $i, j$ (in V), each transmitter $t$ (in $T_i$), and each source-destination pair $(s,d) \in SD$, we introduce the indicator variable $l_{i,j,t}^{s,d}$, defined as:

- $l_{i,j,t}^{s,d} = 1$ if the connection path of from $s$ to $d$ uses *link(i,j,t)*; otherwise $l_{i,j,t}^{s,d} = 0$.

The **objective** of minimizing total power consumer can then be formally expressed as:

$$Min \left( \sum_{i,t,p} p \cdot x_{i,t}^p \right) \quad (1)$$

The objective function (1) is attained subject to certain **constraints**, which are divided into several classes.

### Routing Constraints

To ensure that the (s,d) connection pair is routed correctly, we require, $\forall i \in V$ and $(s,d) \in SD$:

$$\sum_{t \in T_i} \sum_{j \in V} l_{i,j,t}^{(s,d)} - \sum_{t \in T_i} \sum_{j \in V} l_{j,i,t}^{(s,d)} = \begin{cases} +1 & \text{if } s = i \\ -1 & \text{if } d = i \\ 0 & \text{otherwise} \end{cases} \quad (2)$$

To ensure that only a single route is assigned for a given (s,d) pair, we require

$\forall i, j \in V (i \neq j), t \in T_i, (s,d) \in SD:$

$$l_{i,j,t}^{(s,d)} \leq \sum_{(p,\theta_t,\theta_r)} l_{i,j,t}^{p,\theta_t,\theta_r} \cdot g_{i,j,t}^{p,\theta_t,\theta_r} \qquad (3)$$

There are $N \cdot |SD|$ constraints of type (2), and $N \cdot (N-1)|T||SD|/2$ constraints of type (3).

### Delay QoS Constraints

To ensure that the number of hops in the selected route doesn't violate the delay specification, we require

$\forall (s,d) \in SD:$

$$\sum_{(i,j,t)} l_{i,j,t}^{(s,d)} \leq H_{(s,d)} \qquad (4)$$

There are $|SD|$ constraints of type (4).

### Throughput QoS Constraints

To ensure that the throughput specifications are met, we require $\forall i, j \in V (i \neq j), \forall t \in T_i, \forall (s,d) \in SD:$

$$\sum_{(s,d)} l_{i,j,t}^{(s,d)} \cdot Th_{(s,d)} \leq \sum_{(p,\theta_t,\theta_r)} B_{i,j,t}^{p,\theta_t,\theta_r} \cdot (1 - BER_{i,j,t}^{p,\theta_t,\theta_r}) \cdot g_{i,j,t}^{p,\theta_t,\theta_r} \qquad (5)$$

There are $N \cdot (N-1)|T||SD|/2$ constraints of type (5).

### Power Constraints

To ensure that power indicator $x_{i,t}^p = 1$ when transceiver $t$ at node $i$ is transmitting using power $p$, we require that $\forall p \in P, \forall t \in T_i, \forall i \in V:$

$$\sum_{(j,\theta_t,\theta_r)} g_{i,j,t}^{p,\theta_t,\theta_r} \leq N \cdot x_{i,t}^p \qquad (6)$$

where $N$ is the number of nodes, and:

$$\sum_{(j,\theta_t,\theta_r)} g_{i,j,t,p}^{p,\theta_t,\theta_r} \geq x_{i,t}^p \qquad (7)$$

The number of constraints of type (6) and (7) are each $|P||T| \cdot N$.

## Selector Constraints

To ensure that each transceiver is operating at precisely one setting (if at all) $\forall i \in V, \forall t \in T_i$, we introduce selection constraints as follows. If $t$ is an FSO transceiver:

$$\sum_{(j,p,\theta_t,\theta_r)} g_{i,j,t}^{p,\theta_t,\theta_r} \leq 1 \qquad (8)$$

$$\sum_{(j,p,\theta_t,\theta_r)} g_{j,i,t}^{p,\theta_t,\theta_r} \leq 1 \qquad (9)$$

and if $t$ is an RF transceiver, then we require:

$$\sum_{(p,\theta_t,\theta_r)} g_{i,j,t}^{p,\theta_t,\theta_r} \leq 1 \qquad (10)$$

The number of constraints of types (8), (9) and (10) are not more than $2N \cdot |T|$, in total.

## Beam Opening Constraints

To ensure that transceiver $t$ at node $i$ is using the same beam opening during transmission and reception, we require $\forall i, j \in V (i \neq j), t \in T_i, \forall \theta_t \in \Phi$ a constraint of the form:

$$\sum_{(p_x,\theta_t,\theta_r)} g_{i,j,t}^{p_x,\theta_t,\theta_r} + \sum_{(p_y,\theta_m,\theta_n),\theta_n \neq \theta_t} g_{j,i,t}^{p_y,\theta_m,\theta_n} \leq 1 \qquad (11)$$

The number of constraints of type (11) is $N \cdot (N-1)|T||\Phi|/2$.

## Alignment Constraints

To ensure that node transceivers are aligned we require $\forall i, j \in V, \forall t \in T_i, \forall \theta_t \in \Phi,$ (where $\theta_n \neq \theta_t$), a constraint of the form:

$$\sum_{(p,\theta_r)} g_{i,j,t}^{p,\theta_t,\theta_r} + \sum_{(p,\theta_m,\theta_n,l),\theta_n \neq \theta_t} g_{l,i,t}^{p,\theta_m,\theta_n} \leq 1 \qquad (12a)$$

and $\forall i, j, k \in V, \forall t \in T_i, \forall \theta_t \in \Phi,$ where node $k$ is in line of sight of both nodes $i$ and $j$:

$$\sum_{(p,\theta_r)} g_{i,j,t}^{p,\theta_t,\theta_r} + \sum_{(p,\theta_m,\theta_n,l),\theta_n \neq \theta_t} g_{l,k,t}^{p,\theta_m,\theta_n} \leq 1 \qquad (12b)$$

There are at most $N \cdot (N-1)|T||\Phi|/2$ constraints of type (12a) and (12b).

The complexity of any ILP problem depends on the number of variables and constraints in the problem instance. In the above formulation, the factors that determine the number of variables and constraints are the number of nodes ($N$), the number of transceivers at each node ($|T|$), the number of source destination pairs ($|SD|$), the transmission power granularity ($|P|$), and the beam width granularity ($|\Phi|$). The following expression describes the number of variables $W$ involved in the ILP problem:

$$W = N[(N-1)(SDT + P(1+T\theta^2)) + TP] \tag{13}$$

while the number of constraints $W$ involved in the ILP problem is given by:

$$Z = N(N-1)[TSD + 2T + (T-1)\theta] + N[SD + 2PT + 2(T-1)] + SD \tag{14}$$

## IV. LAGRANGIAN RELAXATION

Integer Linear Programming (ILP) is an NP-complete problem, in that its computational complexity is non-polynomial in the number of variables and constraints (unless P=NP). In practice, problem instances quickly become quite large. For example, a scenario with $N=10$, $|SD|=10$, $|T|=4$, $|P|=4$, and $|\Phi|=4$ gives rise to an ILP instance which has *25,740* variables and *5,890* constraints. Finding an optimal solution for such a large ILP problem is computationally infeasible. Thus, in what follows, we consider techniques for finding good (albeit suboptimal) solutions quickly.

In this section, we propose a Lagrangian relaxation (LR) approach that performs constraint relaxations by using Lagrangian multipliers. In the relaxation process, our problem is transformed into a dual problem by adding power constraints (6, 7) and beam opening constraints (11) which are the complicating constraints in the original primal problem.

The **Dual Problem objective** is then

$$\text{Min} \begin{pmatrix} \sum_{i,t,p} p \cdot x_{i,t}^{p} \\ + \lambda_1 ( \sum_{(p_x,\theta_t,\theta_r)} g_{i,j,t}^{p_x,\theta_t,\theta_r} + \sum_{(p_y,\theta_m,\theta_n)} g_{j,i,t}^{p_y,\theta_m,\theta_n} - 1) \\ + \lambda_2 ( \sum_{(j,\theta_t,\theta_r)} g_{i,j,t}^{p,\theta_t,\theta_r} - N x_{i,t}^{p} ) \\ + \lambda_3 ( x_{i,t}^{p} - \sum_{(j,\theta_t,\theta_r)} g_{i,j,t,p}^{p,\theta_t,\theta_r} ) \end{pmatrix},$$

subject to the remaining constraint equations (2-5,8-10,12).

**Updating Lagrangian Multipliers**

Lagrangian multipliers are updated using the procedure outlined in Figure 1, which is based on the LR method [18].

**Feasible Solution Construction**

Despite the fact that Lagrangian Relaxation usually generates infeasible points in the solution space (i.e. violating some constraint), these can be used as a starting point for a "repair strategy" that seeks feasible solutions. We propose a Lagrangian Relaxation with iterative repair heuristic as a mechanism to incrementally modify infeasible points in the solution space into feasible solutions that satisfy all constraints. The strength of the proposed heuristic stems from its simplicity and its propensity, in practice, to achieve a solution that is close to optimal. In our proposed heuristic, we minimize the complexity of the dual problem by deleting unnecessary constraints that don't contain violated variables in their equations. Also, we fix the values of variables that are "not violated"—that is to say, variables that do not appear in any violated constraints. A high level description of the proposed iterative repair heuristic is provided in

Particle Swarm Optimization

While the Lagrangian Relaxation technique described in the previous section is frequently able to provide a feasible (albeit suboptimal) solution for mid-sized problems, it remains computationally prohibitive as problem size increases. To address this, we propose a new approach to solve the topology control problem in Hybrid RF/FSO for large-scale[4] networks. The new approach is based on Particle Swarm Optimization (PSO) technique [23]. The advantage of using the PSO approach over other metaheuristic algorithms is that it is a general technique requiring few parameters to adjust.

---

[4] Large-scale in terms of number of nodes, number of transceivers, and number of source-destination pairs.

## A. Particle Encoding

A particle is encoded as a vector of heterogeneous variables whose values represent a candidate solution of the hybrid RF/FSO topology control problem. These variables are classified into three *types*:

- **X_Power(i,j,t)**: $\forall i, j \in V, \forall t \in T_i$, represents the transmission power from node *i* to node *j* using transceiver *t*.

- **X_Beam(i,j,t)**: $\forall i, j \in V, \forall t \in T_i$, represents the beam opening from node *i* to node *j* using transceiver *t*.

- **X_SD(sd)**: $\forall sd \in SD$, represents a pointer to entries in the route+transceiver assignment enumeration table. This entries in this table are generated by (i) enumerating all the routes generated by the K-shortest paths algorithm for each source-destination pair in *SD*; (ii) for each route generated, enumerate all possible transceiver assignments along the route; and (iii) add each route together with the specific transceiver assignment as an entry in the route+transceiver assignment enumeration table.

## B. Swarm Initialization process

In our model, instead of having the initial population of particles be completely randomly generated, we initialize particles using the first-fit algorithm, as presented in [19]. The set of source-destination pairs *SD* is sorted according to their end-to-end delay requirements, and in case of a equivalent delay requirements the source-destination pair with higher throughput requirements is listed first. The first-fit algorithm simply goes through each source destination pair in order, and searches the route+transceiver assignment enumeration table (starting at the beginning), using the first possible entry that meets the QoS requirements to satisfy the request. If no free entry is available, a random path is selected and the request is marked unsatisfied. To minimize the possibility of being trapped in a local minimum we mixed some totally random particles into the population of first-fit particles. The flow chart in Figure 3 illustrates the swarm initialization process. The variable initialization logic is detailed in the pseudocode shown in Figure 4.

Particle Update Process

Consider a PSO for which $N$ is the number of particles and $M$ is the particle size (i.e. number of variables). Each particle within the PSO moves within the solution space as it attempts to improve its solution quality. This movement occurs according to the particle's position update equation:

$$X_{i,j}^{k+1} = X_{i,j}^{k} + V_{i,j}^{k+1} \qquad (13)$$

Here index variable $i$ ranges from $1$ to $N$, while index $j$ ranges from $1$ to $M$, and $V_{i,j}^{k+1}$ is the updated velocity (i.e. derivative) of variable $j$ within particle $i$, which itself is updated after the $k$th iteration as follows:

$$V_{i,j}^{k+1} = wV_{i,j}^{k} + c1 r1_{i,j}^{k}(P_{i,j}^{k} - X_{i,j}^{k}) + c2 r2_{i,j}^{k}(g_{j}^{k} - X_{i}^{k}) \qquad (14)$$

Equation (14) is referred to as the velocity update equation. Within it, the following quantities are referenced:

$P_{i,j}^{k}$: value of variable j in the best local particle $i$.

$g_{j}^{k}$: the best global value of variable $j$.

$w$: the inertia weight.

$c1$: a positive constant called the cognitive parameter.

$c2$: a positive constant called the social parameter.

$r1_{i,j}^{k}$ and $r2_{i,j}^{k}$: random numbers uniform in [0,1].

C. *Optimization Problem and Fitness Function evaluation*

The constrained topology control problem (first formulated as an ILP and then as a PSO) is transformed into unconstrained problem by defining as the fitness function Fx for our PSO model to be the objective function value plus a penalty $\gamma(x)$ for every unsatisfied constraint $x$. In the literature, the penalty functions considered fall into two categories: stationary and non-stationary. In the former, a fixed penalty is added to the value of the objective function when constraint $x$ is violated. The latter adds considers the penalty as dynamically determined by how far the infeasible point from the constraint. Previous work shows that the results obtained from the non-stationary approaches are closer to optimal than those obtained using stationary approaches [20]. We applying the non-stationary approach to the penalty function, and take our fitness function as follows:

$$Fx = F + h(k)\sum_{i}^{R}\gamma(i) \qquad (15)$$

where $R$ is the number of constraints. We take $h(k) = \sqrt{k}$ (where $k$ is the iteration) in order to increase the penalties for violated constraints as time passes.

**PSO Algorithm**

After defining the swarm structure and the fitness function Fx, the general algorithm of PSO can be applied. The algorithm starts by creating the swarm of particles and assigning each particle its parameters such as the initial position. The algorithm then updates the position of each particle according to velocity and position equations (13,14). In each iteration step, the particle compares its current position with its best-ever position. If the current position turns out to be better, then the current position becomes the new best-ever position. The particle with the best value for the fitness function is chosen to be the swarm's best-particle and the particles in the swarm tend to fly toward this particle. The details are presented as pseudocode in Figure 5.

V. TOPOLOGY CONTROL RESULTS

In this section we provide some experimental results based on the proposed ILP formulation, Lagrangian Relaxation with iterative repair heuristic, and the PSO solution. For our experiments, we assume that the capacity of FSO channel is *500* Mbps, the capacity of RF channel is *50* Mbps, the FSO receiver sensitivity is *-43*dBm, the RF receiver sensitivity is *-84*dBm, and the maximum beam opening is *240* mrad.

A. The first experiment compares ILP versus LR versus PSO solutions for the requested QoS connections *SD* in Table1. The first section of Tables 2 and 3 provides the optimal topology solution we found using the ILP solver. The optimal solution is using the minimum possible power, tuning the precise beam opening, and selecting the accurate channels to meet the joint throughput and end-to-end delay requirements.

One sees from the second sections of Tables 2 and 3 that the Lagrangian Relaxation heuristic provides a sub-optimal solution that is quite close to optimal: LR's total power consumption is 40mW while power consumption was 35mW for the optimal solution obtained by ILP. The third sections of Tables 2 and 3 show that PSO provides a solution that is quite close to the solution obtained by LR: PSO's total power consumption was 45mW, while power consumption was 40mW for the solution obtained by LR.

B. The second experiment investigates the impact of changing source-destination pairs on the generated topology using the LR heuristic. We conducted an experiment similar to the one conducted for omnidirectional wireless networks in [17]. Our conclusions were similar; we found that the network topology in hybrid RF/FSO networks is determined primarily by traffic demand (once node locations are given). Figures 6.a, Figure 6.b, and Figure 6.c demonstrate how the number of links in the topology increases as the traffic demands increase.

C. Next, we investigated how LR reduces the complexity of our ILP instance in terms of reducing the number of constraints (by comparison to the original ILP formulation). To conduct a fair comparison, we conducted our experiment by varying the number of transceivers |T| because this is a common variable between both relaxed constraints and other constraints. In this experiment we computed the percentage improvement in constraints, fixing N=10, |SD|=5, |P|=4, |□□|=4, while varying |T|. The results are shown in Figure 7.

D. Next, we evaluated the quality of the PSO solution for large-scale networks. We placed 30 nodes in a *150m* by *150m* square area. We generated requests between all possible source-destination pairs with uniform probability. Our simulation tool generates *n* requests, and seeks to determine the blocking probability of the network and the total power consumed by all transceivers in the network. The performance of our proposed PSO algorithm operated using 30 swarms with a population size of *150* particles in each swarm, with each PSO simulation running for *1000* iterations.

using LR.

We compared our proposed PSO algorithm with the first-fit heuristic because it has been demonstrated that the first-fit heuristic produces low blocking probabilities [21], and thus serves as a comparison against which to measure our scheme's merits. The proposed PSO heuristic was compared in terms of blocking probability and total power consumption for transmission (over all network nodes).

The first experiment evaluates the impact of increasing the number of connections in the hybrid RF/FSO networks. We fixed *N*=30, *|T|*=5, and varied *|SD|*. Figure 8 shows that the blocking performance of our PSO algorithm is much better than that of the first-fit heuristic under all traffic loads. Also, by looking at the slopes, we conclude that the blocking probability increases as the connection set *SD* increases in size. For example, 10% of the connections did not meet the QoS requirement using the PSO when the *|SD|*=30, but when 35 connections are present, this figure doubles.

The second experiment evaluates on the impact of increasing the number of FSO transceivers on the blocking probability. The network is constructed in the same manner as before. We fixed *N*=30, *|SD|*=50, and varied *|T|*. By looking at the slopes in Figure 9, we see that the blocking probability improves greatly as we increase the number of FSO transceivers.

## VI. Conclusion and future Work

In this paper, we address the topology control problem for hybrid RF/FSO networks. Given a set of end to end delay and throughput quality of service (QoS) requirements, we provided a mathematical formulation of the adaptive control problem, namely the problem of determining how to adjusting transceiver power and beam width to best achieve the QoS requirements. The initial formulation transforms adaptive control problem into an instance of Integer Linear Programming (ILP), which while known to be NP-Complete, served as a starting point and basis for comparison. First, we considered a heuristic solution for ILP, based on Lagrangian Relaxation (LR), using an iterative repair heuristic to find a feasible solution. The LR approach was found to be feasible for mid-sized networks, and produced solutions that were suboptimal yet compared favorably to the optimal solutions computed via much more computationally expensive ILP solver. To extend the model to the domain of large-scale problems, we developed a new approach based on Particle Swarm Optimization (PSO). Our experiments showed that the PSO approach produced solutions that were suboptimal yet compared favorably to the solutions determined by the more expensive LR approach.

Hybrid RF/FSO network is such an environment in which the varying nature of each channel technology can play a major role in the quality of the channel under different weather conditions. In the future, we will extend our model to include an adaptive FEC/ARQ protocol for Hybrid RF/FSO networks. We believe that implementing a hybrid error control system using a combination of FEC/ARQ scheme is considered a promising solution to decide the size of the redundant data based on the channel condition in hybrid RF/FSO networks.

```
For a binary linear minimization problem, with all constraints in canonical form:
1. Begin with each λ at 0, with step size k (problem dependent value)
2. Solve the Lagrangian dual, get solution x.
3. For every constraint violated by x, increase corresponding λ by k.
4. For every constraint with positive slack (rel. to x) decrease corresponding λ by k.
5. If m iterations have passed since the best relaxation value increased, cut k in half.
6. Go to step 2.
```

Figure 1: Baseline Lagrangian Relaxation technique [18] adapted for binary linear minimization problems.

**Definitions**

**allConstraintsMet:** Boolean variable which is true if there is a feasible solution, otherwise it's false.

**violatedVar[]:** Array holds violated variables.
MAXITER: maximum iterations to look for a feasible solution.

**solveLR():** Function that solves the dual problem using Lagrangian relaxation. It returns true if all constraints are met, otherwise it returns false.

**getViolatedvariables():** Function returns violated variables.

**updateDualProblem(violatedVar):** Function updates dual problem based on the provided violated variable from previous Lagrangeian solution.

```
allConstraintsMet=solveLR()
while (allConstraintsMet==false OR i>=MAXITER)
{
    violatedVar = getViolatedvariables()
    updateDualProblem(violatedVar)
    allConstraintsMet = solveLR()
    i = i+1
}
If (allConstraintsMet)
  Print Solution
else
  Print "No Feasible Solution"
```

Figure 2: Lagrangian relaxation with iterative repair heuristic.

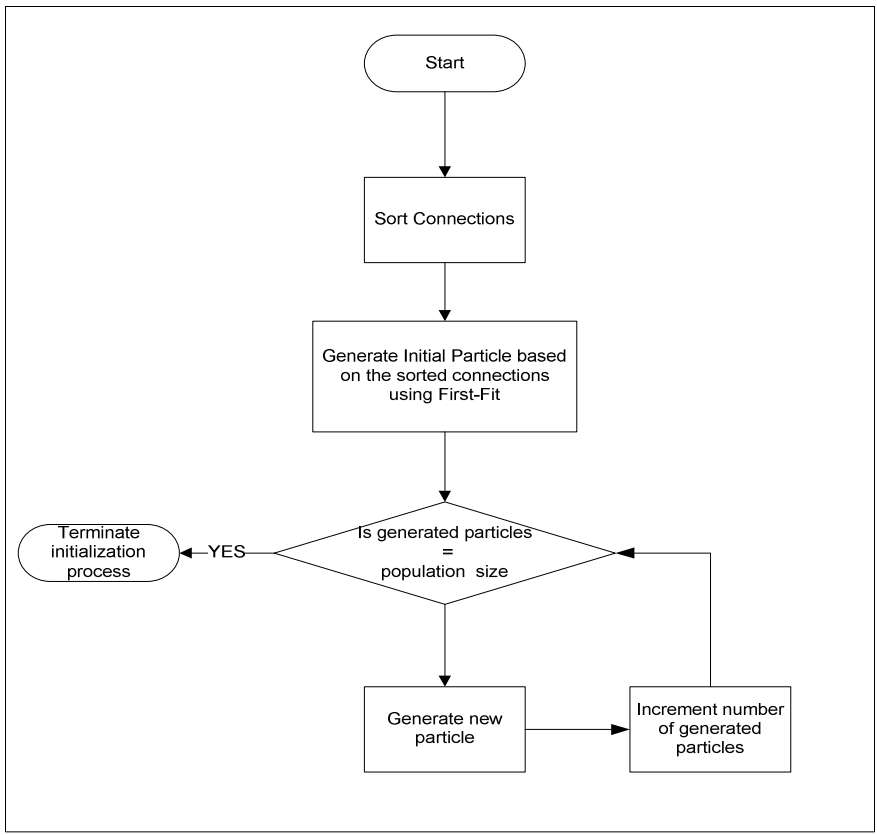

Figure 3: Swarm initialization

```
generateNewParticle: Function that generates new particle
SD: Number of source-destnation pairs
R: Integer value to indicate the percentage of random particles in the initial swarm
ParticleID: The Id of the new partcile that needs to be genrated
generateIntRandom(N): Function geneartes integer number between 0 and N
assignRandomPath: Function assigns a random path index for the source-destination connection
runFirstFit: Function assigns a path index for the source-destination according to the first-fit algorithm
Output: Particle

Particle generateNewParticle(particleId)
{
    if(ParticleID==0)
        for(int sd=0;sd<SD;sd++)
        {
   Particle[i]=runFirstFit(sd);
        }
    else
    {
      int sdCounter=0;
      boolean selectedConnection[SD];
      for(int sd=0;sd<SD;sd++)
       selectedConnection[i]=false;
      while(sdCounter!=SD)
      {
   sd=generateIntRandom(SD);
   if(selectedConnection[sd]==-1)
   {
       if(particleID%R==0)
      Particle=assignRandomPath(sd,ParticleID);
     else
        Particle=runFirstFit(sd,ParticleID);
     selectedConnection[sd]=1;
     sdCounter++;
   }
        }
     }
```

```
   return Particle
}
```

Figure 4: Generating new particle algorithm

```
Do
{
   For each particle
   {
     Calculate the corresponding fitness value
     If the fitness value is better than the
         particle's best fitness value then
     Set the current P vector to the particle's
         current X vector
   }
   Choose the particle with the lowest fitness value
      and make it the global best position

   For each particle
   {
      Calculate the particle's velocity
      Update the particle current position vector X
   }

} while maximum iteration
        or minimum error criteria is not attained
```

Figure 5: PSO algorithm

|   | S | D | Throughput (Mbps) | Delay |
|---|---|---|---|---|
| **1** | 1 | 2 | 5 | 1 |
| **2** | 1 | 5 | 5 | 1 |
| **3** | 2 | 4 | 100 | 2 |
| **4** | 2 | 5 | 100 | 1 |
| **5** | 3 | 1 | 250 | 1 |
| **6** | 4 | 3 | 5 | 1 |
| **7** | 4 | 2 | 5 | 2 |
| **8** | 5 | 4 | 100 | 1 |

Table 1: The QoS Specification

|  | S | D | Route | Selected channels |
|---|---|---|---|---|
| **ILP** | 1 | 2 | 1➔2 | 0 |
|  | 1 | 5 | 1➔5 | 0 |
|  | 2 | 4 | 2➔5➔4 | 2➔1 |
|  | 2 | 5 | 2➔5 | 2 |
|     | 3 | 1 | 3→1 | 1 |
|     | 4 | 3 | 4→3 | 0 |
|     | 4 | 2 | 4→3→2 | 0→3 |
|     | 5 | 4 | 5→4 | 1 |
| LR  | 1 | 2 | 1→2 | 0 |
|     | 1 | 5 | 1→5 | 3 |
|     | 2 | 4 | 2→5→4 | 2→1 |
|     | 2 | 5 | 2→5 | 2 |
|     | 3 | 1 | 3→1 | 1 |
|     | 4 | 3 | 4→3 | 0 |
|     | 4 | 2 | 4→3→2 | 0→3 |
|     | 5 | 4 | 5→4 | 1 |
| PSO | 1 | 2 | 1→2 | 0 |
|     | 1 | 5 | 1→5 | 0 |
|     | 2 | 4 | 2→4 | 1 |
|     | 2 | 5 | 2→5 | 2 |
|     | 3 | 1 | 3→1 | 1 |
|     | 4 | 3 | 4→3 | 0 |
|     | 4 | 2 | 4→2 | 1 |
|     | 5 | 4 | 5→4 | 3 |

Table 2: Routing and channel selection for each connection.

|     | (Link, Transceiver) | Transmitted Power (mW) | Beam Opening (mrad) | Total Power |
| --- | --- | --- | --- | --- |
| ILP | (1→2,0) | 5 | - | 35 |
|     | (1→5,0) | 5 | - | |
|     | (2→5,2) | 5 | 80,80 | |
|     | (3→1,1) | 10 | 80,240 | |
|     | (3→2,3) | 5 | 80,80 | |
|     | (4→3,0) | 5 | - | |
|     | (5→4,1) | 5 | 80,160 | |
| LR  | (1→2,0) | 5 | - | 40 |
|     | (1→5,3) | 5 | 80,80 | |
|     | (1→5,0) | 5 | - | |
|     | (2→5,2) | 5 | 80,80 | |
|     | (3→1,1) | 10 | 80,80 | |
|     | (3→2,3) | 5 | 80,160 | |
|     | (4→3,0) | 5  | -      |    |
|     | (5→4,1) | 5  | 80,160 |    |
|     |        |    |        |    |
| PSO | (1→2,0) | 5  | -      | 45 |
|     | (1→5,0) | 5  | -      |    |
|     | (2→4,1) | 5  | 80,240 |    |
|     | (2→5,2) | 5  | 80,160 |    |
|     | (3→1,1) | 10 | 80,80  |    |
|     | (4→2,1) | 5  | 80,80  |    |
|     | (4→3,0) | 10 | -      |    |
|     | (5→4,3) | 5  | 80,160 |    |

Table 3: Transmitted power and beam opening solution.

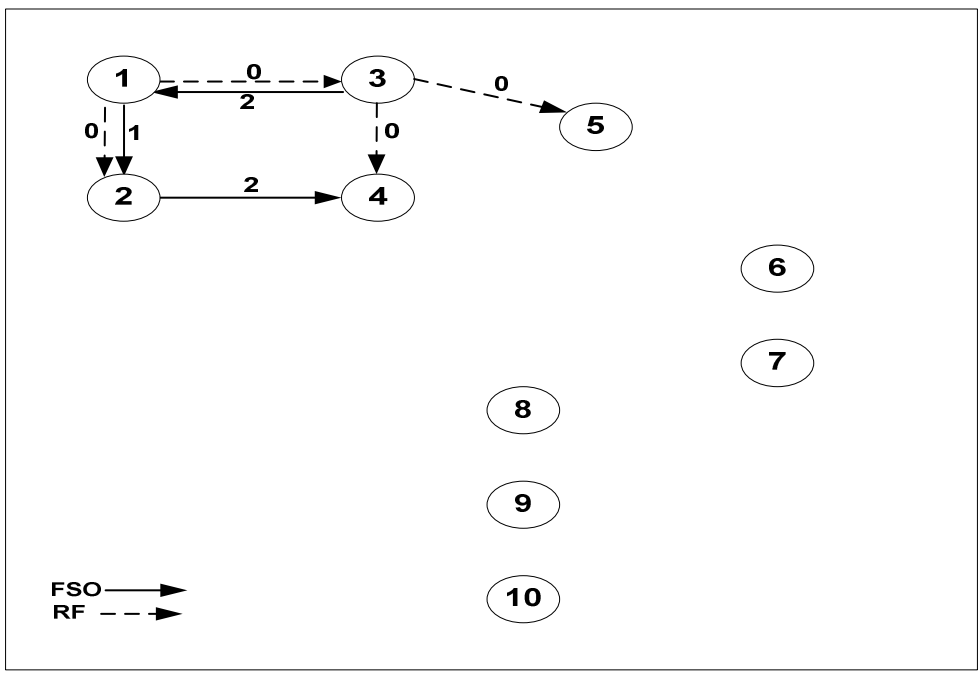

Figure 6.a: network topology for 3 source-destination pairs using LR.

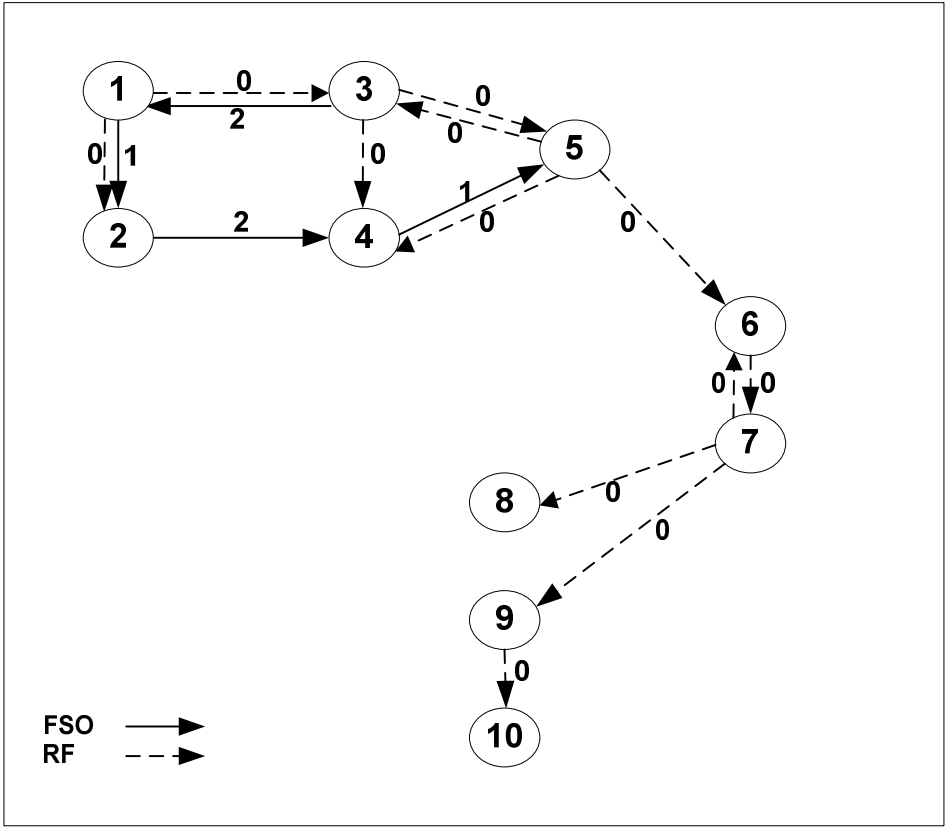

Figure 6.b: The network topology for 5 source-destination pairs using LR.

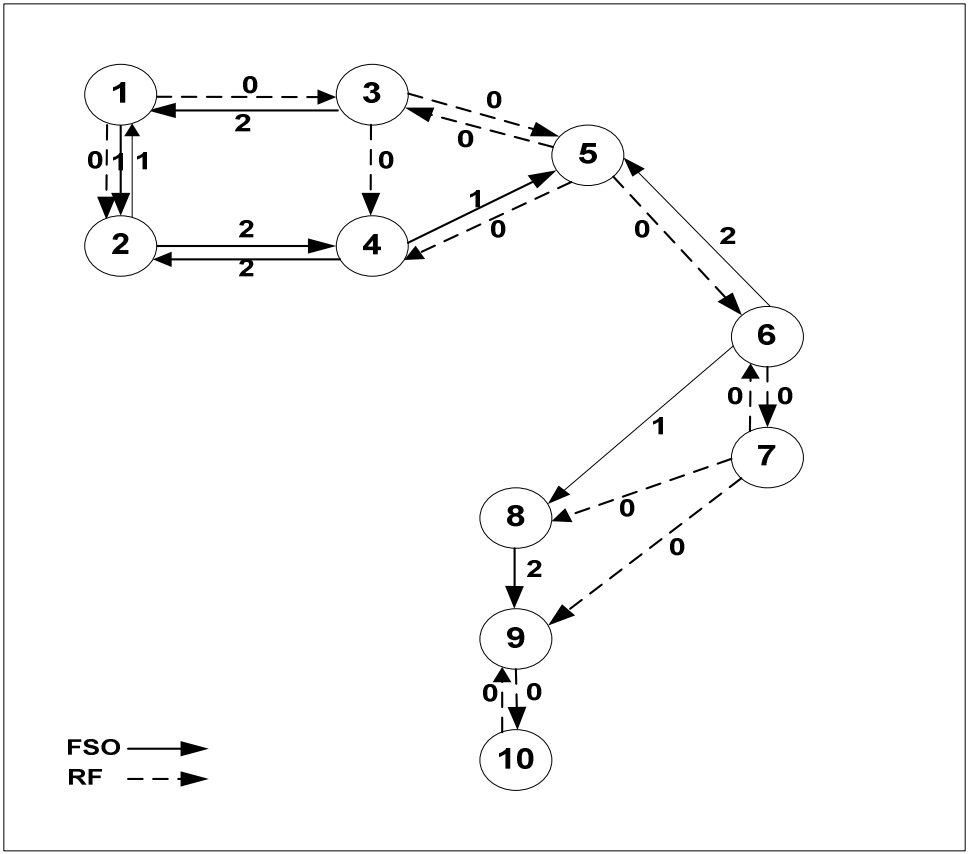

Figure 6.c: The network topology for 10 source-destination pairs

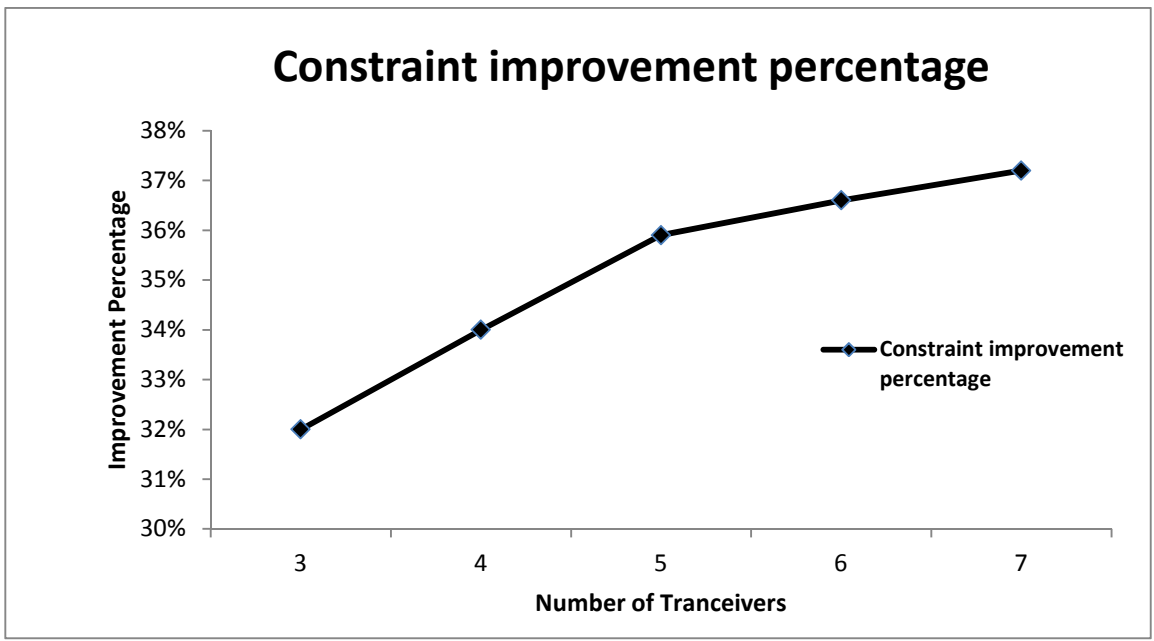

Figure 7: Percentage reduction of constrains using LR.

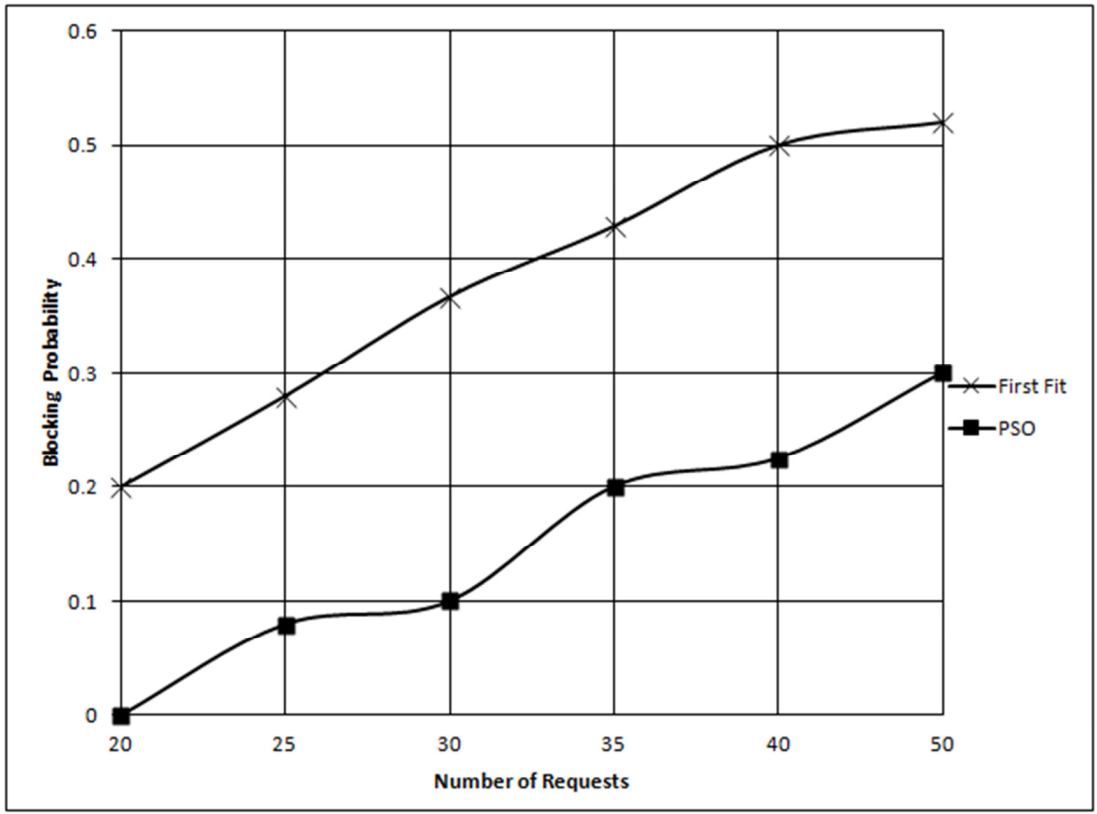

Figure 8: Blocking probability vs. number of connections

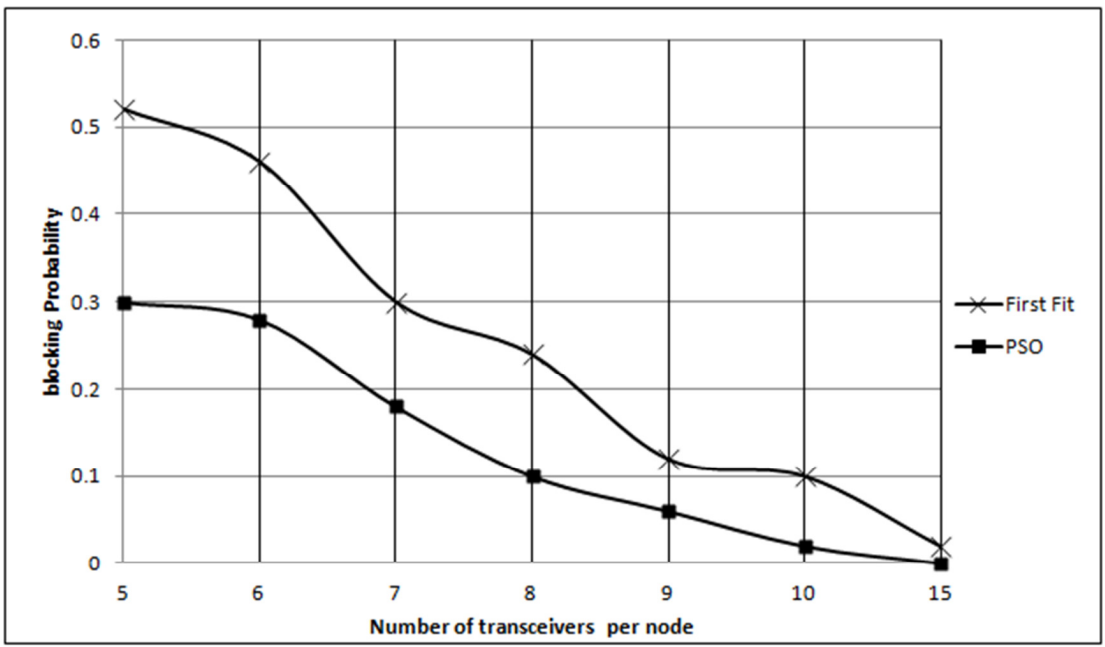

Figure 9: Blocking probability vs. number of transceivers